\documentclass[11pt, a4paper]{article}
\usepackage{amsmath}
\usepackage{typearea}
\pdfoutput=1

\typearea{12}
\title{Gauge invariant observables from
Takahashi-Tanimoto scalar solutions in open string
field theory}
\date{}
\author{Syoji Zeze\footnote{ztaro21@gmail.com}\\ \\
Yokote Seiryo Gakuin High School,\\
147-1 Maeda, Osawa, Yokote, 013-0041 Japan}

\DeclareMathOperator{\erfi}{erfi}
\DeclareMathOperator{\erf}{erf}

\usepackage{graphicx}	
\begin{document}

\maketitle	

\begin{abstract}
Using Maccaferri's formula,  we derive 
new wedge based solutions of open string field theory.
The solutions are gauge equivalent to the Takahashi-Tanimoto scalar
solutions.  The classical action and the gauge invariant overlap are
evaluated analytically. 
We find a perturbative vacuum solution whose gauge invariant observables
vanish.  We also identify a tachyon vacuum solution 
whose gauge invariant observables are identical to those of the Erler-Schnabl solution.
\end{abstract}

\section{Introduction}

One of major motivations for studying string field theory (SFT) is 
to understand its background independence.   In SFT, a string background
specified by a specific conformal field theory (CFT) corresponds to a classical 
solution of the equation of motion.  Therefore, SFT may offer an 
unified framework to derive all possible string vacua from the 
unique action.  One can study relations between different vacua
if corresponding classical solutions are available.

Recently, there has been remarkable
progress \cite{arxiv:1209.4712, arxiv:1307.1203, arxiv:1402.3546} 
for solutions which describe marginally deformed boundary conformal field
theory (BCFT). The novelty of them is 
that they employ the Takahashi-Tanimoto (TT) identity based solution
\cite{hep-th/0202133} as a regularization for the singular OPE between
marginal currents on the boundary.  There, 
the singularities are nicely regularized by an integral of the marginal 
current extended to the bulk.  
In \cite{arxiv:1209.4712}, marginally deformed tachyon vacuum has obtained.   
\cite{arxiv:1307.1203} has derived the gauge invariant overlap (or Ellwood
 invariant) for identity based Takahashi-Tanimoto solution as a
 difference of the overlap between wedge based solutions.  
In \cite{arxiv:1402.3546}, Maccaferri has constructed 
a wedge based solution for marginal deformation. In his work, 
the formula which express a gauge invariant observable 
\footnote{In this paper, we discuss
two well-known gauge invariant observables, i.e.,  the action and
the gauge invariant overlap
\cite{hep-th/0111092, arxiv:0804.1131, arXiv:0804.1541}.
}
as a difference between that of Erler-Schnabl 
solution \cite{arxiv:0906.0979} 
evaluated on the original background and marginally deformed
background plays central role.   Using this formula, 
the gauge invariant observables for
marginal deformation have been evaluated analytically. 
Remarkable feature of \cite{arxiv:1402.3546} is that the
formulas can also be applicable to arbitrary backgrounds as long as a ``seed''
$\Phi$ satisfies the equation of motion. 

As an immediate application of Maccaferri's formula, 
we will consider the TT scalar solution \cite{hep-th/0202133} in this
paper.  As similar to the TT marginal solution, this solution
is identity based.   In sliver frame, the solution is defined by line integrals
of the BRST current $j_B(z)$ and conformal ghost $c(z)$ along 
the imaginary axis convoluted by appropriate functions.  Explicitly, 
\begin{equation}
 \Phi = \int_{-i\infty}^{i \infty}
\frac{dz}{2 \pi i}\left(
f(z) j_{B} (z) + g(z) c(z)
\right) .
\end{equation}
Here $f(z)$ and $g(z)$ is not independent and their relation follows from
the equation of motion.    Much works have been devoted 
\cite{hep-th/0202133, hep-th/0205275, hep-th/0301079, hep-th/0304261, 
hep-th/0405097, hep-th/0502042, hep-th/0506083} to possible choice of $f(z)$,
and it has been conjectured that particular choice give rise to the
tachyon vacuum.  
However, direct derivation of the gauge invariant
observables has not yet available since the
solution is identity based, i.e., it is defined on an infinitesimally
thin strip in sliver frame.   
In this paper, we derive gauge invariant observables, which have been
only studied numerically \cite{hep-th/0302182, hep-th/0502042,
arxiv:0904.1095, arxiv:0910.3026} by extending the TT solution to
a wedge based solution which allows analytic treatment on the world sheet.

This paper is organized as follows.  Section \ref{232526_20Jul14}
introduces the formulas of \cite{arxiv:1402.3546}. In section
\ref{060630_12Jul14}, we apply the formulas to well-known identity based
solutions in $K B c$ subalgebra \cite{arxiv:1008.1104,arxiv:1208.6287}. 
 Although these solutions are well
understood, we present it for demonstration purpose.
Section \ref{233539_20Jul14} deals
with the TT scalar solutions, which is of our main interest.  We
conclude in section \ref{234330_20Jul14} and give discussions.

\noindent \textit{Note added: } While revising this manuscript, we found
two papers  \cite{Kishimoto:2014lua, Ishibashi:2014mua} which deal with
similar subjects with our paper.

\section{Maccaferri's  formulas}
\label{232526_20Jul14}

We begin with a review of 
the formalism developed by Maccaferri \cite{arxiv:1402.3546}, 
on which we heavily rely in this paper.   
The formalism allows us to map an identity based solution
to a wedge based solution whose gauge invariant observable is defined by a 
CFT correlator on a cylinder with nonzero width.  Let $\Phi$ be a ``seed'' solution of the equation of 
motion $Q_B \Phi + \Phi^2 =0$.  
Then, a new wedge based solution is given by,
\begin{equation}
 \Psi = \frac{1}{1+K}\left(\Phi -
 \Phi \frac{B}{1+K'}    \Phi  \right), \label{065354_7Jul14}
\end{equation}
where $K$ and $B$ are familiar elements of the $K B c$ subalgebra and
$K'$ is the deformed generator defined by
\begin{align}
 K' & = Q_{\Phi \Phi} B \notag \\ 
    & \equiv  Q_{B} B +\{B, \Phi   \}.\label{070858_7Jul14}
\end{align}
Note that $K'$ is an element of deformed algebra,
\begin{equation}
 Q_{\Phi \Phi } c = c K' c, \qquad  Q_{\Phi \Phi } B = K', \qquad  Q_{\Phi \Phi } K' = 0,  
\qquad \{c, B\}=1. 
\end{equation}
The solution (\ref{065354_7Jul14}) is gauge equivalent to $\Phi$ since it can be written as a
$\Phi$ dependent gauge transformation
\begin{equation}
 \Psi = (1+ A \Phi) Q_B \left(\frac{1}{1+ A \Phi}\right) + (1+ A \Phi) \Phi \frac{1}{1+ A \Phi},\label{140810_7Jul14}
\end{equation}
where $A=B /(1+K)$.   Notable feature of this prescription is that  gauge invariant observables 
are expressed as a difference between those of the Erler-Schnabl solution \cite{arxiv:0906.0979}
defined on perturbative vacuum and deformed background by $\Phi$, 
respectively.  
For example,  the gauge invariant overlap for closed string vertex
  operator $V$ is given by \footnote{Here $\mathrm{Tr}_V [\Psi] =  
\mathrm{Tr} [ V\Psi]$ and $V = c(i\infty ) c (-i \infty) V_X (i\infty,
  -i\infty)$, where $V_X$ is a weight $(1,1)$ vertex operator in matter
  sector.}
\begin{align}
 \mathrm{Tr}_V [\Psi] & =  \mathrm{Tr}_V [\Psi_{ES, K}] -\mathrm{Tr}_V
 [\Psi_{ES,K'}] \notag \\
      & =\mathrm{Tr}_V 
\left[c\frac{1}{1+K}\right] -\mathrm{Tr}_V \left[c\frac{1}{1+K'}\right],\label{070004_7Jul14}
\end{align}
where $\Psi_{ES,K}$ is the original Erler-Schnabl solution
$\frac{1}{1+K}(c + Q_{B}(B c)   ) $ and
$\Psi_{ES,K'}$ is that defined on the deformed background. 
Quite similarly, the action is also expressed as a difference 
\begin{align}
 S [\Psi] & =  S[\Psi_{ES, K}] -S [\Psi_{ES,K'}] \notag \\
          & =  -\frac{1}{6} \mathrm{Tr} \left[
    \frac{1}{1+K} c   Q_B \left(  \frac{1}{1+K} c    \right)
           \right]
+ \frac{1}{6} \mathrm{Tr} \left[ 
   \frac{1}{1+K'} c  Q_{\Phi \Phi} \left(  \frac{1}{1+K'} c
\right)          
 \right]. \label{072118_7Jul14}
\end{align}
\label{070022_7Jul14} 
As claimed in \cite{arxiv:1402.3546}, derivations of 
(\ref{070004_7Jul14})  and (\ref{072118_7Jul14})  only require
algebraic manipulation in the traces in terms of the $K B c$ subalgebra
and the equation motion of $\Phi$, therefore do not depend on any 
details of $\Phi$. While $\Phi$ can be arbitrary solution, 
 we will consider identity based $\Phi$.

\section{Gauge invariant observables for identity based $K B c$ solutions}
\label{060630_12Jul14}

Next, we apply the Maccaferri's formulas 
to the identity based solutions in $K B c$ subalgebra 
\cite{arxiv:1008.1104,arxiv:1208.6287} to evaluate gauge invariant
observables.  We refer the classification given in \cite{arxiv:1208.6287} where 
the author identified three kinds of gauge orbits, perturbative vacuum,
tachyon vacuum, and the MNT ghost brane.     In following sections, we 
deal with perturbative vacuum and tachyon vacuum which are relevant to
later discussions for the TT solution.

\subsection{Perturbative vacuum}
The ``perturbative vacuum'' solution is the well-known BRST exact string field,
\begin{equation}
 \Phi = \lambda c K B c \quad  (\lambda \neq -1), \label{083917_4Jul14}
\end{equation}
where $\lambda$ is a real constant\footnote{$\lambda= -1$ is excluded
because we cannot construct the similarity transformation discussed in
this section. It also should be noted that this string field
is a part of the MNT ghost brane solution \cite{arxiv:1208.6287} which is not
gauge equivalent to the perturbative vacuum. }. This solution can be
written in pure gauge form $\Phi = U Q_B U^{-1}$ where
\begin{equation}
 U = 1 + \lambda c B , \quad   U^{-1} = 1 -\frac{\lambda}{1+\lambda} c B.
\end{equation}
The deformed generator for this solution is found to be
\begin{equation}
 K' = K + \lambda (c K B + B K c).
\end{equation}
It is not difficult to see that a $n$-th power of $K'$ becomes
\begin{equation}
 K'^n  =   (1+\lambda)^{n-1}
\left\{K^n   + \lambda  (B K^n c + c K^n B)  \right\}. \label{eq:K^n}
\end{equation}
Therefore, a function of $K'$ defined by a power series also satisfies
\begin{equation}
 f(K') = \frac{1}{1+\lambda}
\left\{
f(K_{\lambda}) +\lambda  (B f(K_{\lambda})  c + c f(K_{\lambda}) B)
\right\}\label{083455_4Jul14}
\end{equation}
where $K_{\lambda}  = (1+\lambda) K$.  The formulas (\ref{eq:K^n}) and 
(\ref{083455_4Jul14}) can also
be explained by a gauge transformation $U$ 
since it acts on $K$, $B$ and $c$ in very simple manner,
\begin{equation}
 U K U^{-1}   = \frac{K'}{1+\lambda} ,  \quad
  U c U^{-1}  =  (1+\lambda) c, \quad
  U B U^{-1}  = \frac{B}{1+\lambda}.\label{150935_7Jul14}
\end{equation}

In order to
evaluate the gauge invariant overlap,  we apply (\ref{150935_7Jul14}) to
the last term of (\ref{070004_7Jul14}) to obtain
\begin{align}
 \mathrm{Tr} \left[ V \frac{1}{1+K'} c \right] & =
 \mathrm{Tr} \left[ V U \frac{1}{1+K_{\lambda} } U^{-1}
    c U  U^{-1} \right] \\
& =(1+\lambda )^{-1} \mathrm{Tr} \left[U^{-1}  V U \frac{1}{1+K_{\lambda} } c  \right],\label{192901_7Jul14} 
\end{align}
where $K_\lambda = (1+\lambda) K$. The quantity $U^{-1} V U = U^{-1} c(i
\infty) c (- i \infty) V_{X} U$ can be calculated by using $[c B, c(i
\infty) c(-i \infty)] = c c(-i\infty) -c c(+\infty)$:
\begin{equation}
 U^{-1} V U = V -\frac{\lambda}{1+\lambda} c ( c(-i \infty) -c(i \infty))
  V_{X}.\label{192828_7Jul14} 
\end{equation}
The second term of (\ref{192828_7Jul14}) does not contribute to the
trace (\ref{192901_7Jul14}) because of it collides with another $c$ in the
trace. Therefore the trace reduces to
\begin{equation}
 (1+\lambda)^{-1}  \mathrm{Tr} \left[ \frac{1}{1+K_{\lambda} } c  \right]
= \mathrm{Tr} \left[ \frac{1}{1+K } c  \right]. \label{195136_7Jul14}
\end{equation}
Here the $(1+\lambda)^{-1}$ factor in front of the trace
and the $\lambda$ dependence of
$K_{\lambda}$ in the left hand side of (\ref{195136_7Jul14})
are absorbed into scaling of the correlation function on the cylinder. 
The obtained trace is just the gauge invariant observable for the Erler-Schnabl solution.
This cancels first term of (\ref{070004_7Jul14}), and then the gauge
invariant observable $ \mathrm{Tr}_V [ \Psi ]$  vanishes as expected.

The action can be evaluated similarly.  Using 
(\ref{150935_7Jul14}), the trace in the second term of
(\ref{072118_7Jul14})
is evaluated as
\begin{equation}
 (1+\lambda)^{-3} \mathrm{Tr}\left[
c\frac{1}{1+K_{\lambda} } c K_{\lambda} c
\frac{1}{1+K_{\lambda}}
\right].
\end{equation}
Again, the $1+\lambda$ factors in the trace cancels after scaling
of the correlation function on the cylinder.  Again, this is
exactly same as  the first 
term of (\ref{072118_7Jul14}).   Therefore we obtain vanishing action. 

Although we apply Maccaferri's formulas for a demonstration,
our result can be confirmed by
direct evaluation of the original solution $\Psi$. 
Plugging $\Phi = \lambda c K B c  $ into (\ref{140810_7Jul14}), 
it is soon realized that the entire solution is nothing but a non-real 
form of the Okawa solution \cite{hep-th/0603159}, 
\begin{equation}
 \Psi = F^2 c\frac{K}{1-F^2} B c,\label{205146_7Jul14}
\end{equation}
where  $F^2$ is
\begin{equation}
 F^2 = \frac{\lambda}{1+\lambda}  \times \frac{1}{1+K}.
\end{equation}
The homotopy operator for this solution 
\begin{equation}
 A =\frac{1}{1+\lambda} \frac{1+\lambda + K}{K(1+K)} \label{091353_10Jul14}
\end{equation}
indicates that the BRST cohomology is 
nontrivial due to $1/K$ dependence in (\ref{091353_10Jul14}). 
 Clearly this is a wedge based 
perturbative vacuum solution according to the classification
given in \cite{arxiv:1004.4858}. 

\subsection{Tachyon vacuum}
Next, we consider an identity based solution
\begin{equation}
 \Phi =  c - K c\label{000137_21Jul14}
\end{equation}
which is gauge equivalent to the wedge based tachyon vacuum solution.
Application of the Maccaferri's formulas to (\ref{000137_21Jul14}) 
is straightforward since $K'$
becomes constant as one can easily verify from
(\ref{070858_7Jul14}).  In this case, $K'=1$
and the piece of the solution which is relevant to the gauge invariant
observables is  given by
\begin{equation}
  \frac{1}{1+K'} c  = \frac{1}{2} c.
\end{equation}
The closed string overlap is given by
\begin{equation}
 \mathrm{Tr}_V [\Psi] =  \mathrm{Tr}_V [\Psi_{ES, K}] -
 \frac{1}{2 }\mathrm{Tr}_V [c]. \label{eq:giotach}
\end{equation}
The last term in (\ref{eq:giotach}) is a trace of identity based string field
therefore cannot be evaluated directly.   
 We claim that this trace vanishes as explained below. 
 While there could be several way to
prove this, we follow 
the strategy employed to the case of the marginal deformation
\cite{arxiv:1009.6185, arxiv:1209.4712, arxiv:1307.1203, arxiv:1402.3546}  in which a contribution
from the deformation $\delta K = K'-K$ is expressed as a path ordered
exponential. More precisely, we apply the Schwinger parametrization
\begin{align}
 \frac{1}{1+K'} & = \frac{1}{1+ K+ (-K +1)} \notag \\
  & = \int_{0}^{\infty} dt e^{t (K-1) } e^{-t (1+ K )}.
\end{align}
to the last term of (\ref{eq:giotach}).  We regard $e^{t (K-1)}$ as
deformation for the original background.
In present case,  $\delta K = 1 -K$ commutes with $K$, 
so path ordering is not necessary.  Expanding $e^{t (K-1)}$ in $K$
and regard it as a sum of differentials on $e^{-t (1+K)}$, we have
\begin{equation}
 \frac{1}{1+K'}  = \lim_{s \rightarrow 1}
\int_{0}^{\infty} dt e^{-2 t } e^{-\partial_s} e^{-s t K}. 
\end{equation}
Then, second term of (\ref{eq:giotach}) can be evaluated as
\begin{align}
\int_{0}^{\infty} dt e^{-2 t }
\frac{1}{1+K'} &  = \lim_{s \rightarrow 1}
\int_{0}^{\infty} dt e^{-2 t } e^{-\partial_s}
(s t) \mathrm{Tr}_V[c] \notag \\
& =  \mathrm{Tr}_V[c]  \lim_{s \rightarrow 1}
   (s-1) \int_{0}^{\infty} dt e^{-2 t } t  \notag
 \\
& = 0,\label{192743_22Jul14}
\end{align}
where we have scaled the correlation function on the cylinder in first
line of (\ref{192743_22Jul14}).   Then, 
only the first term of (\ref{eq:giotach}) remains to give a same
value as that of the Erler-Schnabl solution just expected. 

The classical action can be evaluated more straightforwardly by applying
$K' = 1 $ to (\ref{072118_7Jul14}).
In this case, last term in (\ref{072118_7Jul14}) vanishes 
trivially since it is just proportional to $\mathrm{Tr} [c K' c] =  
\mathrm{Tr} [c^3]$\footnote{Although this is identity based, we respect
the algebraic rule $\mathrm{Tr} [c^n]=0$.}. 
Therefore the value of the
classical action also coincides with that of Erler-Schnabl solution.

Again similar to the end of the former section, our result is confirmed
by a form of the whole solution. 
Plugging $\Phi = c - K c $ into  (\ref{140810_7Jul14}), 
one soon realize that the entire solution is nothing but a non-real 
form of the Okawa solution \cite{hep-th/0603159}, 
\begin{equation}
 \Psi = F^2 c\frac{K}{1-F^2} B c,\label{205146_7Jul14}
\end{equation}
where $F^2 = (1-K)/(1+K)$ is just a product of $1-K$ and $1/(1+K)$ which
define two Okawa solutions $c-K c$ and $1/(1+K) c (1+K) B c$.   
According to the classification given in \cite{arXiv:1004.4858}, 
this solution corresponds to tachyon vacuum since homotopy operator
turns out to be $2 B /(1+K)$  which is proportional to that of Erler-Schnabl
solution. Hence BRST cohomology for this solution is trivial.

\section{Takahashi-Tanimoto scalar solution}
\label{233539_20Jul14}

We would like to describe the Takahashi-Tanimoto (TT) scalar
solution \cite{hep-th/0202133} along the line with \cite{1402.3546}.
The solution is written as
\begin{equation}
 \Psi = \int_{-i \infty}^{i \infty} \frac{dz}{2 \pi i}
  \biggl( f(z) j_{B}(z)  + g(z) c(z)  \biggr).\label{070106_8Jul14}
\end{equation}
Here we follow the convention of \cite{arxiv:1402.3546}, where $c(z)$ and
$j_B(z)$ are identity based string fields rather than conformal fields
in sliver frame \cite{hep-th/0511286}.  For example, the string
field $c(z)$ is defined by
\begin{equation}
 c (z) = e^{z K} c e^{-z K}.
\end{equation} 
This description is very useful since $c(z)$
obeys OPE like equation and actually becomes a conformal field once
inserted into the world sheet.  The $z$ integration runs along the
imaginary axis placed at the center of the infinitely thin
vertical strip.  The functions $f(z)$ and
$g(z)$ should be defined on the imaginary axis.  They are not dependent
and fixed by the equation of motion.  In order to solve equation of
motion for (\ref{070106_8Jul14}), we need to evaluate a product of line
integrals.  
As shown in \cite{hep-th/0202133, arxiv:1402.3546}, one of the line
integral in the product can
be converted into a contour integral around other operator in line
integral.  For example, a cross terms of $\int dz f(z) j_{B} (z)$ and $\int
dw g(w) c (w)$ is evaluated as follows:
\begin{align}
 \int_{-i \infty}^{i \infty} \frac{dz}{2 \pi i} f(z)
\int_{-i \infty}^{i \infty} \frac{dw}{2 \pi i} g(w)
\{j_{B}(z), c(w) \}  & =
\oint \frac{dz}{2 \pi i} f(z)
\int_{-i \infty}^{i \infty} \frac{dw}{2 \pi i}
g(w)   j_{B}(z)  c(w) \notag \\
  & =
\oint \frac{dz}{2 \pi i} f(z)
\int_{-i \infty}^{i \infty} \frac{dw}{2 \pi i}
\frac{c(w) \partial c(w)} {z-w} \notag  \\
&=\int_{-i \infty}^{i \infty} \frac{dw}{2 \pi i}
f(w)g(w) c \partial c(w)\label{051305_12Jul14}
\end{align}
In the last line, we use the OPE 
\begin{equation}
 j_{B} (z) c  (w)  \sim \frac{1}{z-w}c \partial c (w).
\end{equation}
Let us introduce a shorthand notation
\begin{equation}
 f \cdot j_{B}  = \int_{-i \infty}^{i \infty} \frac{dz}{2 \pi i}
  f(z) j_{B}(z), \qquad
g \cdot c  = \int_{-i \infty}^{i \infty} \frac{dz}{2 \pi i}
  g(z) c(z).  
\end{equation}
Then, the result (\ref{051305_12Jul14}) is simply stated as
\begin{equation}
\{f\cdot j_{B},  g \cdot c  \} = (f g) \cdot c\partial c.
\end{equation}
In order to derive equation of motion, OPE of between two $j_B$s
\begin{equation}
 j_{B} (z) j_B (w) \sim -\frac{4}{(z-w)^3} c \partial c (w)
- \frac{2}{(z-w)^2} c \partial^2 c (w) 
\end{equation}
is also required. Then another formula follows,
\begin{equation}
\{f_1 \cdot j_{B},  f_2 \cdot j_{B}   \} = 2 (f'_1 f'_2 )  \cdot c\partial c,
\end{equation}
where primes denote $z$ derivative.  $Q_B(  f\cdot j_B )$ and $Q_B
(g\cdot c)$ can also be evaluated by regarding $Q_B \sim 1 \cdot j_B$.  Then 
the right hand side of the equation motion $Q_B \Psi + \Psi^2 = 0 $ is reduced to a line
integral of $c \partial c$ multiplied by 
\begin{equation}
  g(z)  + f(z)g(z)  + f'(z)^2. \label{054748_12Jul14}
\end{equation}
Therefore the equation of motion holds if $f(z)$ and $g(z)$ satisfy
\begin{equation}
 g(z)  + f(z)g(z)  + f'(z)^2 = 0.   \label{094742_2Jun14} 
\end{equation}
For given $f(z)$,  $g(z)$ is easily solved as
\begin{equation}
 g(z) = -\frac{ f'(z)^2 }{1+f(z)}.
\end{equation}
In original TT paper \cite{hep-th/0202133}, solutions for given
$f(z)$ have been considered.
On the other hand,  (\ref{094742_2Jun14}) can also be 
solved for given $g(z)$.  In order to do this, we write $f(z)$ as
\begin{equation}
 f(z) = - \frac{1}{4} \omega(z)^2  -1.\label{130621_23Jul14}
\end{equation}
Then, (\ref{094742_2Jun14}) becomes much simpler
\begin{equation}
 -g(z) +\omega'(z)^2 =0
\end{equation}
which can be easily integrated as
\begin{equation}
 \omega (z) =  \int_{z_0}^{z}  dz' \sqrt{g(z')}.
\end{equation}
Then $f(z)$ is finally given by
\begin{equation}
 f(z) = -\frac{1}{4}  \left(\int_{z_0}^{z}  dz' \sqrt{g(z')}\right)^2 -1.
\end{equation}
Note that $\omega (z)$ is defined up to an integration constant, which
will be fixed later.

Let us now turn to the deformed generator $K'= K + \{B, \Psi\}$, which
is of  our main interest, since it is required for 
the Maccaferri's formulas (\ref{070004_7Jul14}) and
(\ref{072118_7Jul14}) 
in order to evaluate the gauge invariant observables.  From the OPEs
\begin{align}
 j_{B} (z) b (w) & \sim \frac{3}{(z-w)^3} +\frac{j_{g} (w)}{(z-w)^2}
 + \frac{T(w)}{z-w},\\
 b(z)c(w) & \sim \frac{1}{z-w},
\end{align}
one can derive
\begin{align}
 \{h \cdot b, f \cdot j_B \} & = (f h) \cdot  T + ( f'  h) \cdot 
 j_{g} + \frac{3}{2} f''g\label{hb} \\
\{h \cdot b, f \cdot c\}& = f g.\label{091626_10Jul14}
\end{align}
Here a term which is not convoluted with operator denotes a constant
obtained by a line integral.  For example,
\begin{equation}
 f g   = \int_{-i \infty }^{i \infty} \frac{dz}{2 \pi i}
 f(z ) g(z).
\end{equation} 
A sum of (\ref{hb}), (\ref{091626_10Jul14}) for $h(z)=1$ and 
the original $K$ corresponds to $K'$, whose expression is 
\begin{equation}
 K' = (1+f) \cdot T + f' \cdot j_g + \frac{3}{2} f''
+ g.\label{091616_10Jul14}
\end{equation}
It should be noted that $K'$ is well defined if and only if
the constant term $3/2 f'' +g$ is finite.   This impose a constraint
for possible choice of $f(z)$.  

\subsection{Perturbative vacuum}
\label{203431_23Jul14}

Let us first study perturbative vacuum solution.    
The result of section \ref{060630_12Jul14} for perturbative vacuum 
provides a hint to this problem.  As similar to (\ref{150935_7Jul14}),
we expect that (\ref{091616_10Jul14}) is written
as a similarity transformation,
\begin{equation}
 K' = U K U^{-1}. \label{062321_12Jul14} 
\end{equation}
For TT solution, it turns out that the ``twisted'' conformal
transformation realizes this:
\begin{align}
 U & = \exp\left(
\int_{-i \infty}^{i \infty} \frac{dz}{2 \pi i}
 v(z) \tilde{T}(z)
\right) \notag \\
& =e^{v \cdot \tilde{T}}.
\end{align}
Here $\tilde{T}(z)$ is the twisted Virasolo generator \cite{hep-th/0111129, hep-th/0304261}
\begin{equation}
 \tilde{T} (z) = T(z) - \partial j_{g} (z),
\end{equation}
where $j_g (z) = :c (z) b(z): $ is the ghost number current. 
Central charge of the twisted CFT is 24 therefore an OPE between
the energy momentum tensor with other operator
involves an anomalous constant. A finite conformal map generated by $v(z)$
is given by
\begin{equation}
 y(z) = e^{ v(z) \partial} z.
\end{equation}
The details of a proof of (\ref{062321_12Jul14}) is shown
in appendix \ref{090840_10Jul14}. The relation between $f(z)$ 
in (\ref{091616_10Jul14}) and the conformal
transformation $y(z)$ is also given in appendix:
\begin{equation}
 1+ f(y) = \frac{dy}{dz}.  \label{151925_19Jul14}
\end{equation}
This relation arrows us to write a twisted conformal transformation of
$K$ again as a line integral in the new coordinate $y$.  
We require that the conformal map $y(z)$ leaves imaginary axis invariant. 
In terms of a parameter $t$ along the imaginary axis, this means
\begin{equation}
 y( i t)^{*} = - y (it).
\end{equation}
We can further restrict $y$ such that $y(z)^{*} = y(z^{*})$ and 
$y(z) = - y(-z)$.  Then,  $y(z)$ is an odd function under the former
condition.  

 We proceed evaluation of gauge invariant observables.
For the gauge invariant overlap, 
we have
\begin{equation}
 \mathrm{Tr}_V \left[\frac{1}{1+K'} c \right] 
= \mathrm{Tr} \left[U^{-1} V U    \frac{1}{1+K} U^{-1} c U   \right].\label{080121_12Jul14}
\end{equation}
The important aspect of the twisted conformal transformation is that
$c$ transforms as weight 0 tensor.  Therefore, the $c$ insertion in
(\ref{080121_12Jul14})  left invariant,
\begin{align}
 U^{-1} c U  & = U^{-1} c(0) U  \notag\\
            & = c (y^{-1} (0)) \notag \\
            & = c,
\end{align}
by requiring $y(0) = y^{-1} (0) = 0$.   
On the other hand, the transformation of the closed string vertex operator
$V$ involves a conformal factor since transformation of
matter vertex operator $V_X$ is not affected by the
twist. We would like to impose further condition on $y(z)$ which leaves
$V$ invariant.  Since $V$ is located at imaginary infinity, we require
\begin{equation}
 y (\pm i \infty) = \pm i \infty, \qquad y'(\pm \infty)  = 1.
\end{equation}
Under these conditions for $y$, the matter vertex operator $V_X$
produces no conformal factor, and 
the trace is reduced to the Erler-Schnabl's one,
\begin{equation}
 \mathrm{Tr}_V \left[\frac{1}{1+K'} c \right] 
= \mathrm{Tr}_V \left[    \frac{1}{1+K} c   \right].
\end{equation}
As similar to the case of section \ref{060630_12Jul14},
this cancels first term of (\ref{070004_7Jul14}) therefore the gauge invariant overlap vanishes as expected.  
We note that the boundary conditions for $y$,
\begin{equation}
y'(\pm i \infty ) = 1, 
\end{equation}
can be translated into the boundary condition for $f(y)$ according to
(\ref{151925_19Jul14}) as
\begin{equation}
f(\pm i \infty) = 0.
\end{equation}
This is the mid point condition which has been used to define TT solution
\cite{hep-th/0202133, 1402.3546}. 

Evaluation of the action is more straightforward.  The second trace of (\ref{072118_7Jul14})
is 
\begin{equation}
 \mathrm{Tr}\left[
U \frac{1}{1+K} U^{-1} c  U \frac{1}{1+K} U^{-1} c U  K  U^{-1} c
\right] = \mathrm{Tr}\left[ \frac{1}{1+K}  c   \frac{1}{1+K}  c  K   c\right]
\end{equation}
where we use the $U$ invariance of $c$ in right hand side.  This again
cancels the first term of (\ref{072118_7Jul14}) therefore yields
vanishing action. 

As an explicit example of the perturbative vacuum solution, we consider
a normalized Gaussian
\begin{equation}
 f(z) = \frac{1}{\sqrt{2 \pi } s } \exp\left(
{\frac{z^2}{2 s^2}}
\right),
\end{equation}
where $s$ is a real constant.  It is realized that this $f(z)$ and also
$g(z)$ dump well at imaginary infinity.  The anomalous constants in $K'$
, i.e., the last two terms  (\ref{091616_10Jul14})
are also finite.  Therefore we expect that the solution has finite
contraction with wedge based states.  In this example, 
the finite conformal map $y(z)$
is only available as a numerical solution of the differential
equation (\ref{151925_19Jul14}).

\subsection{Tachyon vacuum}
\label{200607_23Jul14}

We next study tachyon vacuum in TT solution.   Let us recall identity
based solutions in $K B c$ subalgebra we discussed in
section \ref{060630_12Jul14}. There,
a crucial difference between tachyon vacuum solution and perturbative
vacuum solution 
is the existence of isolated $c$ term.  Therefore, we consider a
one parameter family such that $g(z)$ in (\ref{070106_8Jul14})
approaches to delta function.  
In such case, the second term of (\ref{070106_8Jul14}) localizes to the
boundary.   We then begin from the normalized Gaussian $g(z)$:
\begin{equation}
 g (z) =  \frac{1}{\sqrt{2 \pi } s } \exp\left(
{\frac{z^2}{2 s^2}}
\right),
\end{equation}
where $s\rightarrow 0$ limit gives delta function.  Corresponding $f(z)$
can be obtained from the formula (\ref{130621_23Jul14}):
\begin{equation}
 \omega (z) = \left( \frac{\pi}{2} \right)^{\frac{1}{4}}
\sqrt{s} \cdot \erfi \left(\frac{z}{2 s}   \right)
\end{equation}
\begin{equation}
  f(z) = - \frac{1}{4}
\left(\frac{\pi}{2} \right)^{\frac{1}{2}}
 s \cdot \left(\erfi \left(\frac{z}{2 s}   \right)\right)^2  -1
\end{equation}
where $\erfi (z) = -i \erf (i z)$ is the imaginary error function. 
An integration constant in $\omega (z)$ is fixed so that $f(z)$ becomes an even
function.   
\begin{figure}[htbp]
\centerline{ \includegraphics[bb=0 0 680 480,scale=0.5]{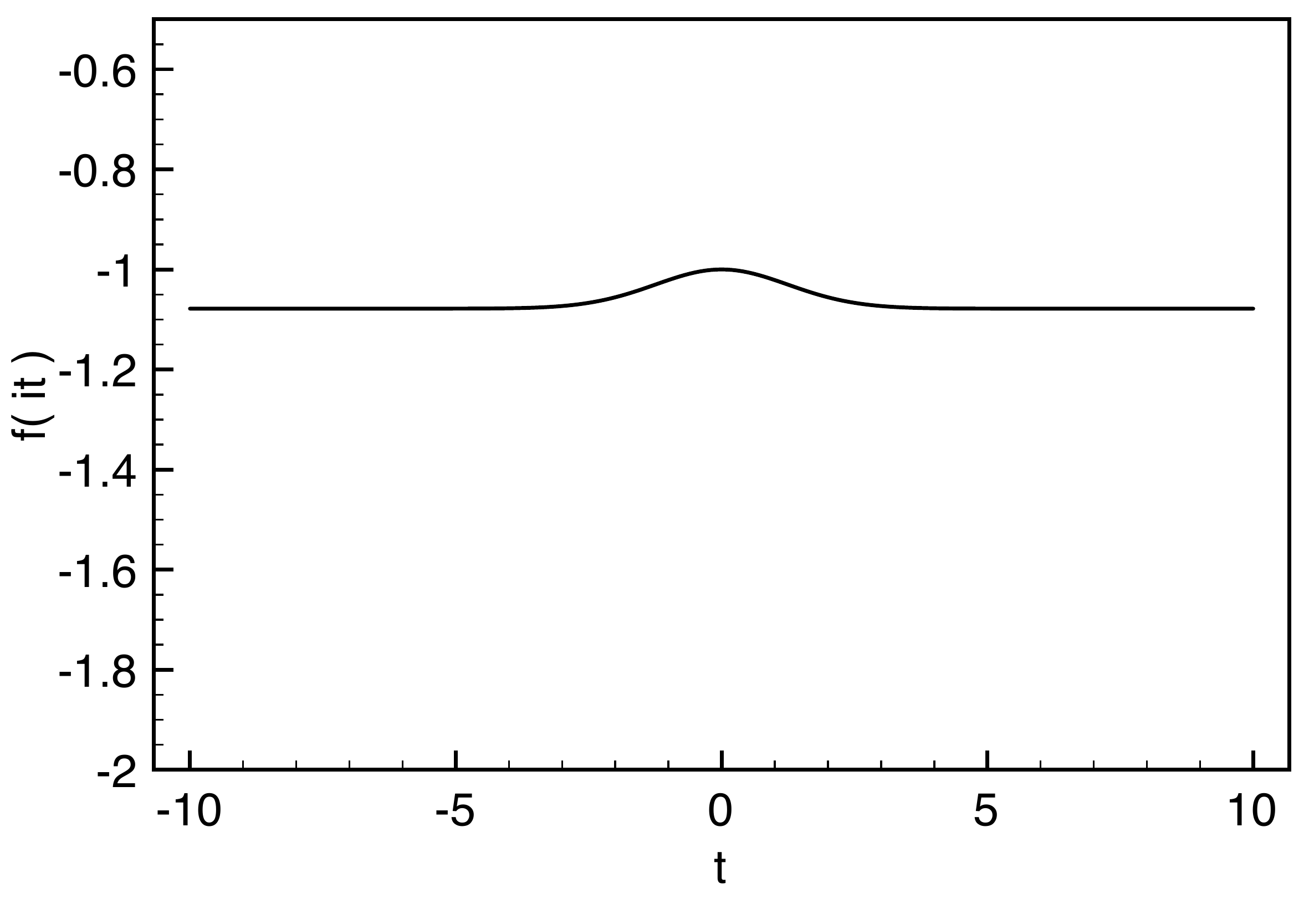} }
\caption{A plot of $f(i t)$ at $s=1.0$. }
\label{141024_23Jul14}
\end{figure}
As seen from fig. \ref{141024_23Jul14}, $f(z)$ is very close to $-1$ and 
there is a small peek around $t=0$.  As $s$ become smaller, the
horizontal line further approaches to $-1$ and the peak become
negligible.  Thus in the $s\rightarrow 0$ limit,
\begin{equation}
 f(z ) \rightarrow -1.
\end{equation}
Therefore, the $(1+f) \cdot T$ terms approaches to zero in this limit. 
On the other hand, the anomalous constant $\frac{3}{2}f''+g$ just
becomes $1$, since $g$ contribution is just an integration of the normalized Gaussian,
and $f''$ contribution becomes zero since it is boundary values of  
 $f'$ which vanishes at imaginary infinity.  Therefore we conclude that 
\begin{equation}
 K'  \rightarrow 1
\end{equation}
in $s\rightarrow 0$ limit.  This limit just corresponds $K'=1$ 
 for the tachyon vacuum solution $c-Kc$.   
Then, gauge invariant observables are
evaluated in exactly same manner as in section \ref{060630_12Jul14}.
This result serves an analytic proof that particular TT scalar solution
corresponds to the tachyon vaccum and successfully derives gauge invariant
observables.

Finally, we note that this example cannot be written in pure gauge
form derived in section \ref{203431_23Jul14}. 
  At least $s \rightarrow 0$  limit corresponds 
to a singular conformal map, since
(\ref{151925_19Jul14}) means  $y'(z) \rightarrow 0$ in this limit, 
therefore the image of $z$ shrinks  to a point.

\section{Summary and discussions}
\label{234330_20Jul14}

We have seen that there is a special limit of TT scalar solution 
in which the deformed generator $K'$ becomes constant.  
In such limit, gauge invariant observables can be evaluated analytically 
with the help of Maccaferri's formula.  We also find a twisted
conformal transformation which maps $K$ to $K'$.  This corresponds 
to the perturbative vacuum.   

Apparently much aspects to be understood.  We have not yet identified
which condition for $f(z)$ distinguish between perturbative vacuum and 
tachyon vacuum.  It is also to be identified that whether the example
given in section \ref{200607_23Jul14} for finite $s$ corresponds
to tachyon vacuum.  It will require a direct evaluation of a trace
which is more complicated than the marginal case.  

Another important feature of the Maccaferri's formalism is the
appearance of the KOS like boundary condition changing (BCC) operator.
In our case of TT scalar, the corresponding BCC operator can be 
written $\sigma_L = \exp ( \chi_h  )$ and 
\begin{equation}
\chi_{h} = \int_{-i\infty}^{i\infty}
\frac{dz}{2 \pi i} h(z)  \varphi (z),
\end{equation}
where $\varphi(z)$ is the bosonized ghost.  Remembering $j_{g} (z) \sim
\partial \varphi (z)$, one can see the formal resemblance of our solution
to boundary deformation.  Our solution looks as if it describes 
a ``boundary'' deformation by the ghost number current. 
Since the ghost current has singular self OPE, regularization by $h(z)$ is
required. 
The tachyon vacuum would be understood as a very
singular limit of such ``deformation''.  
Such interpretation of the tachyon vacuum
by singular deformation will be useful tool,
since it serves ``boundary'' CFT description for the string field theory around
the tachyon vacuum, which might not exist in usual sence.  In such 
context, $\sigma_L$
should be understood as ``boundary removing'' operator.  
However, it is not clear whether above speculation works.
The solution does not look similar to the
KOS solution \cite{1009.6185} since the ghost number current, 
bosonized ghost, and the BCC operator belong to the ghost sector so
the solution behaves quite 
differently from the  marginal solution made from pure matter current.
 It will also be interesting  to apply
the method developed in \cite{arXiv:1406.3021} in order to
render singular ghost current OPE regular one.


\appendix

\section{Proof of (\ref{062321_12Jul14})}
\label{090840_10Jul14}

In this section we prove the formula (\ref{062321_12Jul14}), 
\begin{equation}
U K U^{-1} = (1+f) \cdot T + f' \cdot j_g + \frac{3}{2} f''  -\frac{f'^2}{1+f},\label{092437_23Jul14}
\end{equation}
where $U$ is the twisted conformal transformation:
\begin{equation}
 U = \exp \left(
\oint dz\,  v(z) \tilde{T}(z)
\right).\label{085944_23Jul14}
\end{equation}
In order to evaluate left hand side of (\ref{085944_23Jul14}), we divide
$K$ as
\begin{equation}
 K =\int_{-i \infty}^{i \infty}
\frac{dz}{2 \pi i} T'(z) +
\int_{-i \infty}^{i \infty}
\frac{dz}{2 \pi i} \partial j_g(z) 
\end{equation}
and evaluate each term.  Relevant finite transformations are \cite{Polchinski:1998rq}
\begin{align}
 U \tilde{T}(z) U^{-1}
& = y'^2 \tilde{T} (y)
+\frac{c}{12}
\left[
\frac{y'''}{y'} - \frac{3}{2}
\left(\frac{y''}{y'}\right)^2
\right], \label{091455_23Jul14} \\
 U  \partial j_{g} (z) U^{-1}
& = y''  j_g (y)
+ y'^2  \partial j_g (y)
+\frac{3-2\beta}{2}\left[
\frac{y'''}{y'} - 
\left(\frac{y''}{y'}\right)^2
\right],\label{200154_8Jul14}
\end{align}
where a prime denote derivative with respect to $z$. $c$ is the central
charge and $\beta$ is the parameter that specifies conformal weights of
ghosts.  A sum of (\ref{091455_23Jul14}) and (\ref{200154_8Jul14}) gives a finite
transformation of $T(z)$:
\begin{align}
  U T(z) U^{-1} & = y'^2 \tilde{T} (y) + y''    j_g (y) + y'^2 \partial  j_g (y)
+ \left(\frac{c}{12} +\frac{3-2\beta}{2} \right) \frac{y'''}{y'}
+\left(-\frac{2}{3} \cdot \frac{c}{12} - \frac{3 -2\beta}{2}    \right)
\left(\frac{y''}{y'}\right)^2 \notag \\
& = y'^2 T (y) +  y''  j_g (y) 
+ \frac{3}{2} \frac{y'''}{y'}
-\frac{5}{2}
\left(\frac{y''}{y'}\right)^2.\label{201610_8Jul14} 
\end{align}
We have applied $c=24$, $\beta=2$ and $\tilde{T} (z) = T(z) -\partial
j_g (z)$ in the last line.
Then an integral of
(\ref{201610_8Jul14}) with respect to $z$ gives transformation of
$K$.  
\begin{equation}
U K U^{-1}  = 
\int_{-i \infty}^{i \infty}
\frac{dz}{2 \pi i}  \left(y'^2 T (y) + y''  j_g (y) 
+ \frac{3}{2} \frac{y'''}{y'}
-\frac{5}{2}
\left(\frac{y''}{y'}\right)^2 \right)\label{110422_23Jul14}
\end{equation}
Next, we would like to change variable 
$z$ to $y$ in the integral (\ref{110422_23Jul14}), with assuming that
$y$ leave imaginary axis unchanged.  Then, it turns out that
the first term of (\ref{201610_8Jul14}) matches with $f \cdot T$
in (\ref{092437_23Jul14}) if
we assign \begin{equation}
 y' = 1 + f(y).   \label{093006_23Jul14}
\end{equation}
The constant term of (\ref{201610_8Jul14}) can be written in terms of
$f$ by applying formulas obtained by differentiating (\ref{093006_23Jul14}):
\begin{equation}
 y'' = y' f'(y),  \quad y''' = y'' f'(y) + y'^2 f''(y).
\end{equation}
Then we have
\begin{equation}
U K U^{-1}  = \int_{-i \infty}^{i \infty} \frac{d y}{2 \pi i}  
\left( (1+f(y))  T (y) +  f'(y)   j_g (y) 
+ \frac{3}{2} f''(y) -\frac{f'(y)^2}{1+f(y)}
\right),
\end{equation}
which coincides with the expression for $K'$ given in (\ref{092437_23Jul14}).

\end{document}